\documentclass[preprint,aps,showpacs,prc,nofootinbib]{revtex4-1}
\usepackage{epsfig}

\newcommand{\BV}{\left(\begin{array}{c}}
\newcommand{\EV}{\end{array}\right)}
\newcommand{\BM}{\left(\begin{array}{cc}}

\newcommand{\LQ}{``}
\newcommand{\beqry}{\begin{eqnarray}}
\newcommand{\eeqry}{\end{eqnarray}}
\newcommand{\bqn}{\begin{equation}}
\newcommand{\eqn}{\end{equation}}
\usepackage{color}

\begin{document}

\title{A Modest Revision of the Standard Model}
\author{G.\ J.\ Stephenson,  Jr.}\email{gjs@phys.unm.edu}
\author{T.\ Goldman}\email{tjgoldman@post.harvard.edu}
\affiliation{Dept. of Physics and Astronomy, University of New Mexico, 
	Albuquerque, NM 87501 \\ 
	{\rm and} \\
	Theoretical Division, MS-B283, 
	Los Alamos National Laboratory, Los Alamos, NM 87545}
	
\begin{flushright}
\today\\
{LA-UR-15-21713 (Version 3)}\\
{arXiv:1503.04211v3}\\
\end{flushright}

\begin{abstract}
With a modest revision of the Standard Model, in the fermion mass sector only, the systematics 
of the fermion masses and mixings can be fully described and interpreted as providing 
information on matrix elements of physics beyond the Standard Model. A by-product 
is a reduction of the largest Higgs Yukawa fine structure constant by an order of 
magnitude. The extension to leptons provides for insight on the difference between 
quark mixing and lepton mixing  as evidenced in neutrino oscillations. The substantial 
difference between the scale for up-quark and down-quark masses is not addressed. 
In this approach, improved detail and accuracy of the elements of the current mixing 
matrices can extend our knowledge and understanding of physics beyond the Standard 
Model. 
\end{abstract}

\maketitle

\section{Introduction}

For several decades, the quantum numbers and corresponding gauge interactions 
that distinguish the different generations of fermions have been sought without overt success. 
The various efforts to understand the fermion masses have ranged from substructure 
to Grand Unification. The former approaches include rishons\cite{Harari} and technicolor\cite{Holdom}, 
but always encounter a fundamental problem: For relativistic constituents, the limiting 
gaps between eigenstates tend to a constant value, as in the MIT bag model\cite{Kiskis} 
(surprisingly parallel to the eigen structure for the non-relativistic harmonic oscillator). 
With sufficiently strange potentials (e.g., the \LQ dracula" potential\cite{Kostelecky}), the 
lowest few states may be forced to match the increasing gaps found in the real world, 
but they ultimately tend to a constant gap and predict additional states within the range 
of experiment that remain unseen. The latter  approach has found some interesting 
relations between quarks and leptons\cite{Georgi-Glashow,seesaw} and even some 
mixing angles\cite{Fritzsch} but no convincing overall solution has been obtained. Efforts 
along these lines have continued.\cite{recentarXiv}

We consider a revision of the Standard Model (SM), in the fermion mass sector only, by taking a 
contrary point of view, namely that, within the (revised) SM, all of the fermions with a given 
electric charge should be viewed as having  {\em nothing} that makes their right-chiral, 
weak interaction singlet components distinguishable to the Higgs boson. That is, we take 
the Higgs coupling to be completely insensitive to \LQ generation" and discard the Yukawa 
coefficients that have been (artificially) inserted in the SM to reproduce the observed masses. 

In the past, many others have constructed similar mass matrices by starting from various 
symmetry assumptions. A few of them, of which we are aware, are referenced here.~\cite{others, 
Jarl} They are all \LQ top-down" approaches, making initial symmetry assumptions. Conversely, 
we take a \LQ bottom-up" starting point focusing on the {\em absence} of a known symmetry or 
quantum numbers. This may be viewed as an accidental $S_3$ symmetry but that is not the 
fundamental nature of our basic assumption. Rather, we assume that the final form of the fermion 
mass matrix is determined by all possible (loop) corrections to the SM from physics beyond the 
SM (BSM physics) and invert the relation to extract information on some matrix elements of BSM 
physics. 
This should be contrasted especially with the approach in previous work, such as for example 
in Refs.(\cite{referee}), where the number of unknown BSM parameters is reduced by assuming 
only limited forms for the initial mass matrices. 

We apply this concept here to the quarks and comment on related implications for leptons,  
reserving a full discussion of the leptonic system, including neutrinos and their additional 
complications, to a later paper. This approach amounts to a small change in the mass 
sector of the SM, (hence, a modest revision, albeit a \LQ massive" one) which provides an 
interpretation, in terms of BSM physics, of the deviations from the direct mass implications 
of our revised version of the SM. 

We obtain a consistent description of the quark masses and weak interaction current 
mixing for BSM corrections on the order of 1\%, and constrain relations among the (many) 
allowed BSM parameters, with values that are \LQ natural" in the classic sense ($O(1)$). 
We find that non-Cartan sub-algebra components of the BSM symmetry-breaking corrections 
are required. 

Although this approach proposes a resolution of the differences in fermion masses between 
\LQ generations", it does not attempt to address the differences in mass scale between the 
fermions of different electric charge, generally described as within \LQ families". Our approach 
does, however, reduce the number of these questions to two for the quarks and provides for a 
similar reduction for the leptons. 

\subsection{Mass in the SM}

We first briefly review the structure of SM mass terms that we will revise. They are set by 
(arbitrary) Yukawa couplings of the (one, or in supersymmetry, at least two) Higgs boson(s) 
to the weak interaction active left-chiral doublets and weak interaction \LQ sterile" (except 
for their $U(1)_B$ quantum numbers) singlets. The form is 
\beqry 
{\cal L}_m = & & \Sigma_{i} Y_{Ui} \overline{\left(\frac{1+\gamma_5}{2}\right) \Psi_{Ui} }
<\phi^0 , \phi^+ > \left(\frac{1-\gamma_5}{2}\right) \left[ \begin{array}{c} \Psi_{Ui }
\\ \Psi_{Di} \end{array} \right]  + h.c. \nonumber \\ 
& + & \Sigma_{i} Y_{Di} \overline{\left( \frac{1+\gamma_5}{2}\right) \Psi_{Di} }
<\phi^- , \phi^{0*}> \left(\frac{1-\gamma_5}{2}\right) \left[ \begin{array}{c} \Psi_{Ui}
\\\Psi_{Di} \end{array} \right] + h.c.
\eeqry
where the Higgs may be the same or different in the two sets of terms. 

Unlike the comparable terms for the interaction with the weak gauge bosons, which is 
naturally in the current basis, this Dirac notation is in the mass eigenstate basis and 
suppresses the information that pairs of independent Weyl spinors are involved. In the 
SM, the $Y_{i}$ are taken to equal the mass of the appropriate fermion divided by the 
vacuum expectation value of the Higgs boson. 

In a more general notation, a Dirac bispinor is composed of two Weyl spinors~\cite{vile}
\bqn
\Psi = \left[ \begin{array}{c} \xi \\ \chi \end{array} \right] 
\eqn
where these two chiral spinors are constructed from another pair ($\zeta_a$ and $\zeta_b$) 
with a fixed phase relation, as
\beqry
\xi & = & \zeta_a + \zeta_b  \\
\chi & = & \sigma_2 (-\zeta^*_a + \zeta^*_b)
\eeqry
so that the Dirac mass term appears as 
 \bqn
\overline{\Psi}\Psi  = -( \chi^{\dagger}\xi + \xi^{\dagger}\chi).
\eqn
This makes it clear that the Dirac mass term couples two independent Weyl spinors 
(left- and right-chiral) one of which is weak interaction active and the other sterile as 
described above. 

For completeness, we display the Majorana form for a Weyl spinor and its mass term: 
\bqn
\Psi_{M} = \left[ \begin{array}{c} \xi  \\ - \sigma_2  \xi^*  \end{array} \right]  \; , \;\;
\overline{\Psi_M}\Psi_M  = -( \xi^{\dagger} \sigma_2  \xi^* + \xi^T \sigma_2 \xi) \label{eq:MajM}
\eqn
which will be relevant when we refer to neutrinos later.

\subsection{Basic concept}

The weak iso-singlet produced by the coupling of the Higgs boson to the left-chiral projection 
of the quark (generally, fermion) doublet has no known quantum number besides that of the 
U(1) that completes the electric charges of the fields. Similarly, the right-chiral projections 
have no known other quantum number. Hence we are free to choose bases in the left- and 
right-chiral representation spaces such that the entries to the mass matrix for all of the quark 
(fermion) fields of a given charge are identical. The corresponding mass matrix appears to have 
been first described as \LQ democratic" by Jarlskog~\cite{Jarl}, although, for larger dimensional 
forms, it is known in nuclear and condensed matter physics as the origin of the \LQ pairing gap".  
The eigenstates for such a system consist of one (non-zero energy or) massive state with all other 
(two in the case of interest here) eigenstates being (at zero energy or) massless. This provides for 
a natural starting point consistent with the large gap between the heaviest and the lighter known 
fundamental fermions (of each charged type). We address the question of basis choice below in 
Sec.\ref{sec:mrMS}.

Experimentally, however, the lighter fermions in each grouping are not massless. 
This deviation from zero mass for the lighter fermions can be accommodated by assuming 
small deviations from \LQ democracy", consistent with perturbative corrections from BSM 
physics, which quite naturally allows for small mass eigenstates in the final result. A particular 
constraint on the parameters of the BSM physics is required for one of these to be much 
smaller than the other. Under our assumption, these effects afford a glimpse into the nature 
and structure of the BSM physics by requiring specific relationships between some matrix 
elements (that would be calculable from any given BSM theory). 

The constraints can be discerned by examining the matrix, $U$, that relates the 
BSM-corrected mass eigenstates to the weak interaction eigenstates, for the up-quarks, 
and the corresponding matrix, $V$, for the down-quarks, combined into the form $U V^{\dagger}$ 
that produces the Cabibbo-Kobayashi-Maskawa~\cite{Cab, KM} (CKM) matrix in the form as 
described by the Particle Data Group~\cite{PDG} (PDG). That is, our approach recognizes that 
the Higgs coupling to the \LQ active'', left-chiral, weak iso-doublet fermions is aligned with the 
weak gauge boson coupling, and so is diagonal in that (Weyl spinor) basis.  As stated above, 
however, we take as a given that the right-chiral, weak iso-singlet Weyl spinors all present 
themselves indistinguishably to the weak iso-singlet object formed by combining the Higgs 
doublet with the active fermion doublet. The CKM matrix is produced conventionally by the 
misalignment between the mass and weak eigenstates, but here determined from the two 
separate components, $U$ and $V^{\dagger}$. 

Another feature of the mass eigenstates of the $3\times3$ \LQ democratic" matrix is that the 
eigenvectors for the mass eigenstates can take the form of a tri-bi-maximal (TBM) mixture of the 
original (current) eigenstates. One state must be maximally mixed; for the other two, one has 
a degeneracy choice. However, there is no impediment to an overall TBM choice. While useful 
to simplify calculations, this is not particularly significant for the quarks, as the TBM structure 
cancels out in construction of the CKM matrix. (See Eqs.(\ref{eq:CKMform0},\ref{eq:CKMform}) 
below.)  However, it  does have significant implications for the difference between the CKM and 
the corresponding Pontecorvo-Maki-Nakagawa-Sakata~\cite{PMNS} (PMNS) matrix for mixing 
in the lepton sector. We will comment upon this difference in our conclusions. 

\section{Starting point and positive indications}

The tri-bi-maximal (TBM) matrix
\bqn
TBM  =     \left[ \begin{array}{ccc} 
\frac{1}{\sqrt{6}}   & - \frac{1}{\sqrt{2}}  & \frac{1}{\sqrt{3}}  \\
\frac{1}{\sqrt{6}}   &  \frac{1}{\sqrt{2}}  & \frac{1}{\sqrt{3}} \\
-\frac{2}{\sqrt{6}}  &  0 & \frac{1}{\sqrt{3}} \\
\end{array}
\right]					\label{eq:TBM}
\eqn
diagonalizes the ``democratic" matrix 
\begin{equation}
M_{dem} =   \frac{1}{3} \times
 \left[ \begin{array}{ccc} 
 1 &  1  &  1  \\
1 &  1  &  1  \\
 1 &  1  &  1 \\
 \end{array}
\right]
\end{equation}
to 
\beqry
M_m & = & TBM^{\dagger} \times M_{dem} \times TBM  \nonumber \\
& = &  \left[ \begin{array}{ccc} 
0 &  0  & 0  \\
0  &  0 &  0  \\
0  & 0 &  1\\
 \end{array}
\right]				\label{eq:TBMtrans}
\eeqry
where we have chosen the overall scale so the nonzero eigenvalue is unity. 

The efficacy of the \LQ democratic" conjecture can be tested by inverting the 
TBM transformation on the known quark masses (taken from the PDG\cite{PDG} 
and) placed into diagonal mass matrices, {\it viz.},
\bqn
m_u   =  \left[
\begin{array}{ccc}
2.3 & 0 & 0 \\
0 &  1275 &0 \\
0 &  0 & 173500 
\end{array} \right]  \\ 
\eqn
and 
\bqn
 m_d  =  \left[
\begin{array}{ccc}
3.8 & 0 & 0 \\
0 &  95 &0 \\
0 &  0 & 4150 
\end{array} \right]  
\eqn
where all values are expressed in MeV/$c^2$. (We will ignore the significant 
uncertainties and variation with scale of these masses\cite{scale} as the ratios 
vary less dramatically, although the values of even the ratios are not known to 
very high accuracy.) 

We now transform these inversely using the $TBM$ matrix given in 
Eq.(\ref{eq:TBM}) and find that the resulting mass matrices are indeed 
quite \LQ democratic":
\beqry
m_{u-TBM} & = & TBM \times m_{u} \times TBM^{\dagger} \nonumber \\
  & = & (173500) \times  \left[
\begin{array}{ccc}
0.33701 & 0.32966 & 0.33333 \\
0.32966 &  0.33701 & 0.33333 \\
0.33333 &  0.33333 & 0.33334
\end{array} \right]   \label{demoup}
\eeqry
and similarly 
\bqn m_{d-TBM}  =   (4150) \times  \left[
\begin{array}{ccc}
0.34493 & 0.32204 & 0.33303 \\
0.32204 &  0.34493 & 0.33303 \\
0.33303 &  0.33303 & 0.33394
\end{array} \right]  				\label{demodn}
\eqn
where we have scaled out the overall factor of the largest mass in each case. 
Although the true accuracy is, of course, far less, we keep the extra digits 
to display which matrix elements are not identical after the (inverse) TBM 
transformation and so convey the patterns that will survive even substantial 
(within experimental uncertainties) changes in the ratios of the diagonal values. 

This demonstrates that only perturbatively small corrections to a democratic 
starting point are needed. (We have ignored $CP$-violation considerations here, 
but will return to them below.) The deviations from ``democracy" are exceptionally 
small, less than 1\% in the up-quark sector and less than 4\% in the down-quark 
sector (for positive and negative deviations from an average).  It is clear from this 
that something close in structure to $M_{dem}$ (times an overall mass scale, $m$) 
is a reasonable ansatz to consider for an initial mass matrix. (A similar result holds 
for the charged leptons.)

This result confirms that the wide range of quark masses is well described by an 
almost \LQ democratic" mass matrix for each charge set of quarks, leaving only the 
overall scale difference between up-quarks and down-quarks (and also leptons) 
to be understood. We do not address that difference here. 

\section{A Modestly Revised Mass Sector} \label{sec:mrMS}

With these comments and results in mind, we propose the mrSM (modestly 
revised Standard Model) which differs from the SM only in the fermion mass 
sector. In terms parallel to those of the SM used above, we have
\beqry 
{\cal L}_{mr} = & & Y_{U} \Sigma_{i,j} \overline{\left(\frac{1+\gamma_5}{2}\right) \Psi_{Ui} }
<\phi^0 , \phi^+ > \left(\frac{1-\gamma_5}{2}\right) \left[ \begin{array}{c} \Psi_{Uj }
\\ \Psi_{Dj} \end{array} \right]  + h.c. \nonumber \\ 
& + & Y_{D} \Sigma_{i,j} \overline{\left( \frac{1+\gamma_5}{2}\right) \Psi_{Di} }
<\phi^- , \phi^{0*}> \left(\frac{1-\gamma_5}{2}\right) \left[ \begin{array}{c} \Psi_{Uj}
\\\Psi_{Dj} \end{array} \right] + h.c. 
\eeqry
There is now only one $Y$ each for all of the up- and down-quarks and these 
two values are approximately one third of the value of the two largest $Y_{i}$ 
in the SM. 

Of course, as long as each up-quark is related to a particular down-quark by the weak 
interaction, one may choose a \LQ rotation" of the pair to any basis. We reiterate the 
basic points: Because there are no known quantum number restraints, we are free to 
rotate the bases for both the left- and right-chiral fields independently to ensure that the 
mass matrix is democratic. (With some particular BSM theory, this should be the basis 
required by quantum number constraints.) In that basis, it is clear that BSM corrections 
are perturbatively small, a situation in physics always much to be desired. This is 
especially so here, since effects of BSM physics, if they exist at all, must be suppressed. 

We note, as {\it lagniappe}, that this factor of $3$ reduction in these remaining two 
Yukawa couplings to the Higgs field from the largest value in each case in the SM 
improves the formally perturbative character of this part of the SM by reducing the 
(set of) Higgs fine structure constant(s), $Y^2$/$4\pi$, by an order of magnitude. 
Of course, the total size of the effects of quark loops is unaltered, as the sum over 
all channels produces the same significant net effect as in the diagonal mass basis. 
However, the size of the total contribution becomes due to the number of diagrams 
contributing, not to any individual large one. 

We take the full Higgs plus BSM-loop-corrected mass matrix to have the form 
\bqn
{\mathcal M_{mrSM}} = m\times[ {\mathcal M_{dem}}+\epsilon {\mathcal M_{BSM}} ]
\eqn
now in the {\em current} quark basis consistently defined by the Higgs and weak 
vector boson couplings. That is, we define the mass matrix for each set of 3 quarks 
of a given electric charge as $m$ (an overall scale which is approximately one-third 
of the mass of the most massive of each triple of the fermions of a given non-zero 
electric charge) times the matrix  ${\mathcal M}$, where  
\beqry
 {\mathcal M} & = &  \frac{(1+\epsilon \xi)}{3} \times
 \left[ \begin{array}{ccc} 
 1 &  1  &  1  \\
1 &  1  &  1  \\
 1 &  1  &  1 \\
 \end{array}
\right]   \nonumber  \\  
& + & \epsilon \times \left[ \begin{array}{ccc} 
  \sqrt{\frac{2}{3}}y_0+y_3+\frac{1}{\sqrt{3}}y_8  &  y_1-I{y_2} & y_4-I{y_5} \\
y_1+I{y_2} & \sqrt{\frac{2}{3}}y_0-y_3+\frac{1}{\sqrt{3}}y_8 & y_6-I{y_7} \\
y_4+I{y_5}  & y_6+I{y_7} & \sqrt{\frac{2}{3}}y_0-\frac{2}{\sqrt{3}}y_8 \\
 \end{array}
\right]               \label{eq:genmtx}
\eeqry

This allows for the most general set of perturbative (for small $\epsilon$) deviations 
possible for a Hermitean $3\times 3$ matrix from the democratic mass matrix produced 
by uniform Higgs' coupling in each quark charge sector. The coefficients are chosen 
to match the normalization of the standard Gell-Mann $SU(3)$ ($U(3)$) basis matrices. 

The BSM corrections are all taken to be proportional to the small quantity,  $\epsilon$, 
defined by the diagonal matrix of known mass eigenvalues, (again, with the overall 
scale factored out)
\begin{equation}
m \times  
 \left[ \begin{array}{ccc} 
 \epsilon \delta &  0  & 0  \\
0  &  \epsilon &  0  \\
0  & 0 &  1\\
 \end{array}
\right]	 \label{eq:diag}
\end{equation}

The overall factor $(1+\epsilon \xi)$ in Eq.(\ref{eq:genmtx}) is introduced to rescale the largest
eigenvalue of ${\mathcal M}$ to unity, needed to account for the effect of the $\epsilon y_0$ 
correction from the BSM physics.  We will see below how a correlation between some of these 
$y_i$ constants (which we view as matrix elements of BSM physics loop corrections, see below) 
reduces the smallest eigenvalue from $\epsilon$ to $ \epsilon \delta $. 

As is apparent from Eq.(\ref{eq:diag}), $\delta$ is the ratio of the lightest mass of the three quarks 
(with the same electric charge) to the mass of the intermediate mass quark, and $\epsilon$ is the 
ratio of that quark to the most massive of the three. In particular, for the quark mass values referred 
to above, 
\beqry
\epsilon_u =   7.35 \times 10^{-3},   \,\,   & & \,\,   \delta_u =   1.8 \times 10^{-3}   \nonumber \\
\epsilon_d =    2.29 \times 10^{-2},   \,\,  & &  \,\,    \delta_d =   4.0 \times 10^{-2}
\eeqry 
Even the largest of these values easily qualifies as a small expansion parameter. We will see below 
that the $\delta$s do not significantly influence our results, so the largest perturbation is provided by 
$ \epsilon_d$. 

The quantities $y_0$, $y_3$ and $y_8$ describe a subset of the possibilities for symmetry breaking 
from the BSM physics: While $(1+\epsilon \xi)$ is only an overall scale revision,  $y_3$ and $y_8$ 
parameterize the usual Cartan sub-algebra of $SU(3)$ (dynamical) symmetry breaking allowed for 
3 complex degrees of freedom, and $y_0$ parameterizes the correction allowed by the $U(1)$ factor 
of the overall $U(3)$. These are all multiplied by a factor of $\epsilon$ in the expectation that the BSM 
corrections are small, as is suggested by the data referred to above. It is apparent from the numerics 
above that this assumption is self-consistent. 

We also allow for off-diagonal corrections, reflecting a possible more complete breaking 
of the (accidental)  $U(3)$ symmetry. We have included these elements of necessity, having 
found that, with only $y_3$ and $y_8$, the desired result of the CKM matrix (in the form 
of the combination $U V^{\dagger}$)  cannot match all of the (moduli of the) elements of 
that matrix as reported by the PDG~\cite{PDG}. However, as emphasized by Leviatan~\cite{Ami}, 
some elements of a partially broken symmetry may survive when additional components 
outside the Cartan subalgebra, or even outside the parent spectrum generating algebra, 
appear.  Interestingly, the converse does not hold. We will see below that it is possible to 
describe the mass and current misalignment with at least one and possibly both $y_3$ and 
$y_8$ vanishing as long as components outside the Cartan subalgebra contribute.

\section{Explicit construction} 

After TBM transformation as in Eq.(\ref{eq:TBMtrans}),  an intermediate form of the mass matrix with BSM 
corrections as given in Eq.(\ref{eq:genmtx}) becomes
\beqry
& & {\mathcal M}_{int}    =   \left[ \begin{array}{ccc} 
0 & 0  & 0  \\
0 & 0  & 0  \\
0 & 0  & 1+ \epsilon \xi   \\   \label{eq:Mintz}
\end{array}
\right]  + \epsilon \times      \\
& & \mkern-36mu    \left[ \begin{array}{ccc} 
\sqrt{\frac{2}{3}} y_0 - \sqrt{\frac{1}{3}}y_8  +\frac{1}{3} ( y_1 - 2 y_{4p6}) &
-\frac{1}{\sqrt{3}}( y_3 -  y_{4m6} + I (y_2 - y_{5m7}))&
 \sqrt{\frac{2}{3}} y_8 +\frac{1}{3\sqrt{2}}( 2y_1 - y_{4p6}) - \frac{1}{\sqrt{2}} I y_{5p7}     \\
-\frac{1}{\sqrt{3}}( y_3 -  y_{4m6}-I (y_2 - y_{5m7}))  & \sqrt{\frac{2}{3}} y_0+ \frac{1}{\sqrt{3}}y_8 -  y_1  & 
-\frac{1}{\sqrt{6}} (2y_3 + y_{4m6} -  I (2y_2 +y_{5m7})) \\
 \sqrt{\frac{2}{3}} y_8 +\frac{1}{3\sqrt{2}}( 2y_1 - y_{4p6}) + \frac{1}{\sqrt{2}}  I y_{5p7} & 
 -\frac{1}{\sqrt{6}} (2y_3 + y_{4m6} + I (2y_2 +y_{5m7}) &   \sqrt{\frac{2}{3}} y_0+\frac{2}{3}( y_1 +  y_{4p6}) \\
\end{array}
\right] 	\nonumber		
\eeqry

We have introduced a $p$ and $m$ simplifying notation:  $y_4-y_6 = y_{4m6}$, 
$y_4+y_6 = y_{4p6}$, $y_5-y_7 = y_{5m7}$ and $y_5+y_7 = y_{5p7}$ to indicate 
sums and differences of the numerically labelled $y_i$'s. This combines them into 
shorter labelled forms to reduce the complexity of the appearance of the form of 
${\mathcal M}_{int}$. We now reformulate Eq.(\ref{eq:Mintz}) in terms of phases as 
\beqry
& & {\mathcal M}_{\phi}    =    \left[ \begin{array}{ccc} 
0 & 0  & 0  \\
0 & 0  & 0  \\
0 & 0  & 1 
\end{array}
\right] + \epsilon \times   \nonumber     \\
& &   \left[ \begin{array}{ccc} 
\sqrt{\frac{2}{3}} y_0 - \frac{1}{\sqrt{3}}y_8  +\frac{1}{3} ( y_1 - 2 y_{4p6}) &- y_a e^{-\imath \alpha} & y_b e^{\imath \beta}   \\
-y_a e^{\imath \alpha} & \sqrt{\frac{2}{3}} y_0+ \frac{1}{\sqrt{3}}y_8 -  y_1  & -y_c e^{-\imath \gamma} \\
y_b e^{-\imath \beta} & -y_c e^{\imath \gamma}  &  \xi +  \sqrt{\frac{2}{3}} y_0+\frac{2}{3}( y_1 +  y_{4p6}) \\
\end{array}
\right] 		\label{eq:Mintphi}
\eeqry
where 
\beqry
y_a  &    =   &  \frac{1}{\sqrt{3}} \sqrt{ (y_3-y_{4m6})^2 +  (y_2-y_{5m7})^2 } \label{eq:ya} \\
\alpha  &    =   & - {\rm arctan}\left(\frac{y_2-y_{5m7}}{y_3-y_{4m6}}\right) \\
y_b  &    =   &  \frac{1}{3\sqrt{2}} \sqrt{ (2\sqrt{3}y_8+2y_1-y_{4p6})^2 + 9 (y_{5p7})^2 }  \label{eq:yb}  \\
\beta  &    =   & {\rm arctan}\left(\frac{ 3 y_{5p7}}{2\sqrt{3}y_8+2 y_1-y_{4p6}}\right) \\
y_c  &    =   &   \frac{1}{\sqrt{6}} \sqrt{ (2y_3+y_{4m6})^2 +  (2y_2+y_{5m7})^2 }  \label{eq:yc}   \\
\gamma  &    =   &  {\rm arctan}\left(\frac{ 2y_2+y_{5m7}}{2y_3+y_{4m6}}\right)
\eeqry
and we take the positive values of the square roots. 

There is a phase freedom associated with each of the three fermions or eigenvectors. Although the overall 
phase can have no effect, we can make use of two phases to transform the mass matrix to the form 
\beqry
& & {\mathcal M}_{\zeta}    =    \left[ \begin{array}{ccc} 
0 & 0  & 0  \\
0 & 0  & 0  \\
0 & 0  & 1 
\end{array}
\right] + \epsilon \times   \nonumber    \\
& &   \left[ \begin{array}{ccc} 
\sqrt{\frac{2}{3}} y_0 - \frac{1}{\sqrt{3}}y_8  +\frac{1}{3} ( y_1 - 2 y_{4p6}) & -y_a & y_b \; e^{\imath \zeta}   \\
-y_a  &  \sqrt{\frac{2}{3}} y_0+ \frac{1}{\sqrt{3}}y_8 -  y_1  & -y_c  \\
y_b  \; e^{-\imath \zeta} & -y_c  &  \xi +  \sqrt{\frac{2}{3}} y_0+\frac{2}{3}( y_1 +  y_{4p6}) \\
\end{array}
\right] 		\label{eq:zetaphs}
\eeqry
where 
\bqn 
\zeta =  \alpha +  \beta +  \gamma  \label{eq:zeta} 
\eqn
is the only physically meaningful quantity and $y_a$, $y_b$ and $y_c$ are unaltered. 

We have come this far by transforming the mass matrix by the TBM matrix as in Eq.(\ref{eq:TBMtrans}), 
but we need further transformations to carry out the diagonalization. Here, we only display this 
calculation through order $\epsilon$. We proceed in two steps, first block diagonalizing with a generic 
matrix, $X_{3\rightarrow2}$, which will apply for both up-quarks and down-quarks with appropriate 
parameter values. We then complete the calculation by diagonalizing the remaining $2\times2$ 
block with the generic matrix, $X_{2x2}$. The total transformation is given by 
\bqn 
X_{tot}  = TBM \times X_{3\rightarrow2} \times X_{2x2}    \label{eq:tot}
\eqn
We use the notation, $X$, to reflect the fact that the form of the matrices that produce the 
diagonalization of the (approximate) mass matrix will be applied to both the up-quarks, where the 
full matrix that produces diagonalization is conventionally labelled as $U$, and to the down-quarks, 
conventionally labelled as $V$. 

\subsection{Block diagonalization}

We now first block-diagonalize the $3\times3$ matrix in Eq.(\ref{eq:zetaphs}). With the expectation 
that the third component will dominate the eigenvector for the large (unit) eigenvalue of the 
full system, we choose an eigenvector of the form 
\bqn
vec{3} = \left[ \begin{array}{c}
 \alpha \\
 \beta \\ 
 1\\
\end{array}
\right]
\eqn
where we expect $\alpha$ and $\beta$ to be of order $\epsilon$, and solve the 
first two of the equations in
\bqn
{\mathcal M}_{\zeta} \; vec{3} = 1 \cdot vec{3}   \label{eq:vec3phs}
\eqn
for  $\alpha$ and $\beta$. Their values to leading order in $\epsilon$ are 
\beqry
\alpha & =  &   \epsilon \, y_b  \, e^{\imath \zeta} \\  \label{alfa3phs}
\beta  & =  &  -\epsilon \, y_c   \label{beta3phs}
\eeqry 
The value of $\xi$ is determined by the requirement that the trace of ${\mathcal M}_{\zeta}$ 
equal the trace of the scaled matrix in Eq.(\ref{eq:diag}): 
\bqn
\xi = 1 + \delta - \sqrt{6} y_0
\eqn
Then the value of $y_0$ can be found from the third component relation in Eq.(\ref{eq:vec3phs}).
To the leading order approximation in $\epsilon$ used above, it is 
\bqn
y_0  =  \frac{1}{2\sqrt{6}}(3(1+\delta) + 2 (y_1 + y_{4p6}) ) \label{eq:y0val}
\eqn

It is now straightforward to construct two vectors orthogonal to $vec3$. With these, 
and after normalizing all three vectors, we can block diagonalize ${\mathcal M}_{\zeta}$ 
as an intermediate step to the full diagonalization. Since we are only working to order 
$\epsilon$, we do this in a way that minimally affects the $2\times2$ subspace, choosing 
\beqry
vec{1} & =  & \left[ \begin{array}{c}
{1} \\
0 \\
-\alpha^* 
\end{array}
\right]
 \\ 
vec{2} & =  & \left[ \begin{array}{c}  
0 \\
1 \\
-\beta 
\end{array}
\right]
\eeqry 
The minus signs appear because the full matrix of eigenvectors  is necessarily of the form of 
a complex rotation and $\beta$ is real, so it does not require complex conjugation. 

Together with $vec3$, the two vectors $vec1 $ and $vec2$, allow us to present 
an explicit unitary transformation matrix, $X_{3\rightarrow2}$, that produces, through order $\epsilon$, 
a block-diagonalized form of ${\mathcal M}_{\zeta}$, namely
\bqn
X_{3\rightarrow2}   =   \left[ \begin{array}{ccc} 
1 & 0  &  \epsilon \, y_b \, e^{\imath \zeta} \\
0 & 1  & -\epsilon \, y_c  \\
- \epsilon \, y_b \, e^{-\imath \zeta} &  \epsilon \, y_c    & 1 \\
\end{array}
\right]
\eqn
which block diagonalizes ${\mathcal M}_{\zeta}$ to the desired unit eigenvalue and a 2-by-2 block 
\bqn
 M_{2x2}   =   \epsilon  \times 
 \left[ \begin{array}{cc}
 \frac{1}{2}(1+\delta) -\frac{1}{\sqrt{3}}y_8 +\frac{1}{3} (2y_1 - y_{4p6} )  &   -y_a     \\
 - y_a  &  \frac{1}{2}(1+\delta) +\frac{1}{\sqrt{3}}y_8 -\frac{1}{3} (2y_1 - y_{4p6})  \\
\end{array}
\right] 
\eqn
(The first two entries of the third row and column of the full matrix are zero to order $\epsilon^2$ by construction; 
the values of $\xi$ and $y_0$ above may be used to confirm the unit value of the resulting $(3,3)$ 
element.) 

Note that, due to the isolation of the phase to the $(1,3)$ element of ${\mathcal M}_{\zeta}$, 
there are no order $\epsilon$ imaginary terms in $M_{2x2}$.  It is here that the advantage 
of avoiding an order $\epsilon$ imaginary contribution in $M_{2x2}$ becomes apparent: 
Once the factor of $\epsilon$ is removed from  $M_{2x2}$, it is clear that $X_{2x2}$, 
the matrix that diagonalizes $M_{2x2}$, would include an eigenvector with an order one 
imaginary phase which would lead to a large $CP$ violation, inconsistent with observation, 
and so require a delicate cancellation to occur. The isolation of the net phase to the $(1,3)$ 
element of ${\mathcal M}_{\zeta}$ enforces this necessary result automatically. 
(A parallel result obtains if the phase freedom is used to combine the separate phases in 
${\mathcal M}_{\phi}$ into the invariant total phase $\zeta$, but in the $(2,3)$ element of 
${\mathcal M}_{\zeta}$ instead.) 

\subsection{Completion of Diagonalization}

To complete the analysis and proceed to apply constraints on the BSM parameters, we need 
to diagonalize $M_{2x2}$. This is, of course, straightforward; we simplify the notation by 
identifying
\bqn
y_f  = \frac{1}{\sqrt{3}}y_8 - \frac{1}{3} (2y_1 - y_{4p6} ) \label{eq:yfval}
\eqn
so that the eigenvalues are   
\bqn
 \frac{1}{2}(1+\delta) \pm \sqrt{y_a^2 + y_f^2}
\eqn
Next we define $\omega$ by 
\bqn
{\rm tan}(2\omega) = -\frac{y_a}{y_f}
\eqn
and set  
\beqry
y_a  &  = &    \left[  \frac{1- \delta}{2}\right] {\rm sin}(2\omega)   \label{eq:yaomega} \\
y_f   &  = &    \left[  \frac{1- \delta}{2}\right]{\rm cos}(2\omega)   \label{eq:yf} 
\eeqry
so that the required values of the eigenvalues, $\delta \epsilon$ and $\epsilon$, are obtained. 
The $3x3$ matrix that diagonalizes the $2x2$ block in the full mass matrix is then just 
\bqn
X_{\omega}   =   \left[ \begin{array}{ccc}
 {\rm cos}(\omega) &  {\rm sin}(\omega)   & 0  \\
  -{\rm sin}(\omega)  & {\rm cos}(\omega) & 0  \\
   0  & 0 & 1 
\end{array}
\right] 
\eqn
where we have relabeled $X_{2x2}$ as $X_{\omega}$ for notational simplicity. The full 
matrix that completely diagonalizes the mass matrix to order $\epsilon^2$ is now
\bqn
X_{tot} =  TBM \times X_{3\rightarrow2} \times X_{\omega}
\eqn
as we described in Eq.(\ref{eq:tot}). Explicitly, 
\bqn
X_{tot}   =   \left[ \begin{array}{ccc}
 {\rm cos}(\omega) &  {\rm sin}(\omega)   & \epsilon y_b \, e^{\imath \zeta}  \\
  -{\rm sin}(\omega)  & {\rm cos}(\omega) & - \epsilon y_c  \\
   - \epsilon (y_b \, e^{\imath \zeta} {\rm cos}(\omega) + y_c  {\rm sin}(\omega)) & 
   - \epsilon (y_b \, e^{\imath \zeta} {\rm sin}(\omega) - y_c  {\rm cos}(\omega)) & 1 
\end{array}
\right]     \label{resultX}
\eqn

While every particular BSM physics model will specify the values of the $y_i$ that enter into 
the determination of $\omega$, it should be clear from the above that inversion of the mass 
spectrum is not sufficient to determine a value for $\omega$. Thus, it remains an effective 
free parameter in what follows. 

\section{Fitting to the CKM matrix}  \label{fitz}

To compute the CKM matrix, we need the result in Eq.(\ref{resultX}) evaluated for both the 
up-quarks, and for the down-quarks. Again, the separate matrices for these are conventionally 
labelled $U$ and $V$ respectively~\cite{PDG}, so that
\bqn
CKM = U \times V^{\dagger}.  \label{CKMdefn}
\eqn

However, we reiterate that the PDG description is one in which these matrices transform from 
mass eigenstates to current eigenstates, but our derivation above is for the transformation of 
current eigenstates to mass eigenstates. Hence, the Hermitian conjugates are interchanged 
and the $V^{\dagger}$ of the PDG is our $X_{tot}$ for the down-quarks and similarly, 
their $U$ is the Hermitian conjugate, $X_{tot}^{\dagger}$,  of our $X_{tot}$ for the up-quarks. 

On combining the results for up-quarks and down-quarks to produce the equivalent of the CKM 
matrix, 
\bqn
CKM =  X_{\omega_u}^{\dagger} \; X_{3\rightarrow2_u}^{\dagger} \; TBM^{\dagger} \; TBM \; 
                                    X_{3\rightarrow2_d} \; X_{\omega_d}  \label{eq:CKMform0}
\eqn
we see that the $TBM$ factor cancels out in the product. Hence, it is sufficient to calculate 
\bqn
CKM =  X_{\omega_u}^{\dagger} \times X_{3\rightarrow2_u}^{\dagger} \times 
                               X_{3\rightarrow2_d} \times X_{\omega_d}  \label{eq:CKMform}
\eqn

Continuing to work only at first order in the small quantities, we define
\beqry
 A_{du}  & = &   \epsilon_{d} y_{bd} e^{I\zeta_d} -\epsilon_{u} y_{bu} e^{I\zeta_u}  \label{eq:CKMparamA}  \\
 B_{du}  & = &   -(   \epsilon_{d} y_{cd} - \epsilon_{u} y_{cu} ) \label{eq:CKMparamB}
\eeqry
and again make use of phase freedoms:  In the CKM matrix, there are six available, three each from 
the up-quark and down-quark sectors. As before, one overall phase can have no effect, but four of 
the remaining five can be chosen to obtain
\beqry
CKM(1,3)  & = &  [{\rm cos}(\omega_u)  A_{du}  -  {\rm sin}(\omega_u) B_{du}] e^{-I\Phi} \nonumber \\
CKM(2,3)  & = & [{\rm sin}(\omega_u)  A_{du}    +  {\rm cos}(\omega_u)  B_{du}] e^{-I\Phi} \nonumber \\
CKM(3,1)  & = & [- {\rm cos}(\omega_d)  A_{du}^{\star}   + {\rm sin}(\omega_d) B_{du}] e^{I\Phi} \nonumber \\
CKM(3,2)  & = &[ -{\rm sin}(\omega_d) A_{du}^{\star}  -  {\rm cos}(\omega_d)  B_{du}] e^{I\Phi} \label{eq:allCKMparams}
\eeqry
where $\Phi$ is the last of the five that can be freely chosen. We chose a value for this so that 
the imaginary part of $CKM(2,3)$ vanishes, corresponding most closely to the convention of 
the PDG\cite{PDG} for the $CKM$ matrix:
\bqn
\Phi  = {\rm arctan}\left(\frac{{\rm sin}(\omega_u) Im(A_{du})}
{{\rm cos}(\omega_u) B_{du}+{\rm sin}(\omega_u) Re(A_{du})}\right).
\eqn
so that 
\beqry
 {\rm sin}(\Phi) &  = & \frac{{\rm sin}(\omega_u) Im(A_{du})}
{\sqrt{[{\rm cos}(\omega_u) B_{du}+{\rm sin}(\omega_u) Re(A_{du})]^2+[{\rm sin}(\omega_u) Im(A_{du})]^2}} \\
&  = & \frac{{\rm sin}(\omega_u) Im(A_{du})}
{\sqrt{{\rm cos}^2(\omega_u) B_{du}^2+{\rm sin}^2(\omega_u) A_{du}^{\star}A_{du} 
+{\rm sin}(2\omega_u) B_{du} Re(A_{du})}} \nonumber \;\;\; {\rm and} \\
{\rm cos}(\Phi) &  = & \frac{{\rm cos}(\omega_u) B_{du}+{\rm sin}(\omega_u) Re(A_{du})}
{\sqrt{[{\rm cos}(\omega_u) B_{du}+{\rm sin}(\omega_u) Re(A_{du})]^2+[{\rm sin}(\omega_u) Im(A_{du})]^2}} \\
&  = & \frac{{\rm cos}(\omega_u) B_{du}+{\rm sin}(\omega_u) Re(A_{du})}
{\sqrt{{\rm cos}^2(\omega_u) B_{du}^2+{\rm sin}^2(\omega_u) A_{du}^{\star}A_{du} 
+{\rm sin}(2\omega_u) B_{du} Re(A_{du})}} \nonumber 
\eeqry
which leaves 
\bqn
CKM(2,3) = {\sqrt{[{\rm cos}(\omega_u) B_{du}+{\rm sin}(\omega_u) Re(A_{du})]^2+[{\rm sin}(\omega_u) Im(A_{du})]^2}}
\eqn
and accomplishes the desired result. 

To leading order in $\epsilon$, $CKM(3,3) = 1$, and the relation between $\omega_u$ and 
$\omega_d$ is immediately fixed by the ($2 \times 2$) Cabibbo rotation in the light quark sector 
 (the $CKM(i,j)$ entries for $i=1,2$ and $j=1,2$):
\beqry
CKM_{C} & =  &  \left[ \begin{array}{cc}
 {\rm cos}( \omega_d - \omega_u) &  {\rm sin}( \omega_d - \omega_u)     \\
- {\rm sin}( \omega_d - \omega_u)   &  {\rm cos}( \omega_d - \omega_u)
\end{array}
\right]  \nonumber \\
& =  &  \left[ \begin{array}{cc}
 {\rm cos}( \Theta_C) &  {\rm sin}( \Theta_C)     \\
- {\rm sin}(\Theta_C)  & {\rm cos}( \Theta_C) 
\end{array}
\right] 
\eeqry
(plus order $\epsilon$ corrections), {\it i.e.,} we identify
\bqn
\Theta_C = \omega_d - \omega_u \;  \label{eq:thetaC} .
\eqn
At this point, it is apparent that the $CKM$ mixing depends solely on the difference 
between the diagonalizations of the up-quarks and the down-quarks, as it must. 

\subsection{PDG evaluation}

The PDG~\cite{PDG} provides only the absolute values of the entries of the $CKM$ matrix. 
It also presents a matrix form that has only real entries in the first row and third column of 
the $CKM$ matrix, except for the $(1,3)$ matrix element. This is achieved by locating the 
one required phase in the matrix that produces rotation about the second axis, where the 
sequence of rotations is first about the third axis (almost Cabibbo), next about the second 
axis, and finally about the first axis, proceeding from right to left in the products in the usual 
way, {\it viz.}
\beqry
CKM_{PDG} & =  &  \left[ \begin{array}{ccc}
1 &  0   & 0  \\ 0   & c23 & s23 \\ 0 & -s23   & c23 \\ \end{array} \right] 
\times 
 \left[ \begin{array}{ccc}
c13 &  0   & s13\, e^{-\imath \chi}  \\  0   &1 & 0 \\ -s13\, e^{\imath \chi} & 0   & c13 \\ \end{array} \right] 
\times 
 \left[ \begin{array}{ccc}
c12 &  s12   & 0  \\ -s12   & c12 & 0 \\ 0 & 0   & 1  \\ \end{array} \right]   \nonumber \\
& =  & \left[ \begin{array}{ccc}
c13\,c12 &  c13\,s12   & s13\, e^{-\imath \chi}  \\ 
-c13\,s12 -c12\,s23\,s13\, e^{\imath \chi}   & c23\,c12 -s12\,s23\,s13\, e^{\imath \chi}  & s23\,c13 \\ 
-c12\, s13\, e^{\imath \chi} -c12\,c23\,s13\, e^{\imath \chi}  & -s23\,c12 -s12\,c23\,s13\, e^{\imath \chi}  & c23\,c13 
\end{array} \right] 
\eeqry
where as usual, $c13 = {\rm cos}(\theta_2)$, etc. and we have changed the PDG phase 
notation from $\delta$ to $\chi$ to avoid confusion with our mass ratio parameter above. 

We follow the PDG structure precisely below and our construction above agrees with 
its structure through first order in $\epsilon$ as the sines of all of the angles are small 
(see below). However, we need to have explicit imaginary components for all of the 
matrix entries, rather than only moduli as reported by the PDG. To proceed, we have 
constructed a version of the PDG result where we assume that all three of the mixing 
angles reside in the first quadrant. This is not justified, but demonstrates how the constraints 
on BSM parameters may be extracted were such information available. 

Taking values from the PDG~\cite{PDG}, and using their parametrization, we obtain 
central values for the real and imaginary parts of these quantities in the relations: 
 \beqry
CKM13  & = & 0.001067 - 0.003386 I  \nonumber  \\ 
CKM23  & = &  0.04141 + 0.0 I \nonumber \\
CKM31  & = & 0.008375 - 0.003294 I \nonumber  \\
CKM32  & = & -0.04057 - 0.0007690 I  \label{eq:CKMelvals}
\eeqry
where the {\it rhs} in each case is the corresponding entry of the matrix $CKM_{PDG}$ 
when, as noted above,  the particular set of signs for the sines is chosen corresponding 
to all three angles  being in the first quadrant. Also, using the entries in the upper left 
$2\times2$ block, (which are real through first order in small quantities as $s13$ and 
$s23$ are both small) we estimate the value of the Cabibbo angle, $\Theta_C$, as
\bqn
 \Theta_C = 0.2291  \label{eq:thetaCval} 
\eqn
{\it i.e.,} approximately $13.2^{o}$. 

Other choices for extracting the full matrix elements could be investigated as well, but this 
demonstrates that at least one solution exists. We have investigated a number of alternatives and 
find that the largest differences, apart from signs, are in the real and imaginary parts of $CKM13$ 
and part of $CKM32$, but the changes are not large, e.g., $\sim 20$\%. However, if the phase is 
placed in an alternate location, for example so that the first row entries are {\it all} 
real, then larger changes are obtained in the real and imaginary parts, although the moduli are 
maintained, of course. 

We uniformly present 4-digit values for consistency, but the changes between the 2012 and 2014 
reports suggest that in a number of cases the values are not known to better than two digits, at most, 
although some of the uncertainties are a small fraction of a percent. Due to the larger uncertainties, 
however, we conclude that carrying out our analysis to order $\epsilon^2$ is not warranted at this 
time.  

\subsection{CP Violation}

We have taken advantage of the phase freedoms, particularly as noted by Kobayashi and Maskawa 
(KM)~\cite{KM}, and used them to ensure that a large $CP$-violation, which would be inconsistent 
with experiment, does not arise within the  ($2 \times 2$) light quark sector.  We must, however, 
examine what $CP$-violating implications are introduced by the BSM parameters that we have 
introduced that produce complex amplitudes.  In particular, we can examine whether this is sufficient 
to be the only source of $CP$-violation. 

The invariant characterization of $CP$-violation was described by Jarlskog~\cite{CPV}. The Jarlskog 
invariant quantity, which we label ${\mathcal J}$, appears only at order $\epsilon^2$, and is given 
by~\cite{PDG}
\beqry
{\mathcal J}  & = & {\mathcal I m} [ CKM_{i,j} CKM_{k,l} CKM^*_{i,l} CKM^*_{k,j} ]\nonumber \\
& = & {\rm cos}(\theta_{12}) {\rm sin}(\theta_{12}) {\rm cos}(\theta_{23}) {\rm sin}(\theta_{23}) 
{\rm sin}(\theta_{13}) {\rm cos}^2(\theta_{13}) {\rm sin}(\delta) \nonumber \\
& = &  (3.06 \pm 0.21) \times 10^{-5}
\eeqry
up to an overall sign ambiguity, in the standard PDG representation of the CKM matrix. 

At the first order in $\epsilon$ level of approximation, only the combination of matrix elements 
[$i=j=2,k=l=3$] reproduces the correct result for ${\mathcal J}$:
\beqry
{\mathcal J_{2233}} & = &  \pm
{\mathcal Im} [ CKM(2,2) CKM(3,3) CKM^*(2,3) CKM^*(3,2) ]\nonumber \\
 & = &   \pm {\rm cos}(\Theta_C) \; 1 \;  {\mathcal Im}\{[{\rm sin}(\omega_u)  A_{du}    
 		 +  {\rm cos}(\omega_u)  B_{du}] [{\rm sin}(\omega_d) A_{du}^{\star}   
		 +  {\rm cos}(\omega_d)  B_{du}] \}	\nonumber \\
 & = & \pm  {\rm cos}(\Theta_C)  {\rm sin}(\Theta_C) B_{du}  {\mathcal Im} (A_{du}) \label{eq:J2233form}
\eeqry
 We have checked that completing the unitary structure of $X_{tot}$ to second order in $\epsilon$ 
 reproduces the correct result from any $i,j,k,l$ combination. With the value of ${\mathcal J}$ known, 
 this provides one constraint on one pair of the combined parameters. 

\subsection{BSM parameter constraints}

The parameter combinations in Eqs.(\ref{eq:allCKMparams},\ref{eq:CKMparamA},\ref{eq:CKMparamB},\ref{eq:thetaC}) 
display the fact that there are 4 quantities that can be directly related to the $CKM$ matrix elements: 
$Re(A_{ud})$, $Im(A_{ud})$, $B_{ud}$ and either one of $\omega_u$ or $\omega_d$. Using the 
values of the $CKM$ matrix elements in Eqs.(\ref{eq:CKMelvals}), we can extract values for these 
combinations of BSM matrix elements. 

As seen above, $\Theta_C = \omega_d - \omega_u$. If we further simplify the notation by defining 
\beqry
X \;\;\; = \;\;\;  B_{du}   & = &  -(\epsilon_d y_{cd} -  \epsilon_u y_{cu})  \nonumber \\
Y = Re(A_{du}) & = &  \epsilon_d y_{bd}{\rm cos}(\zeta_d) - \epsilon_u y_{bu} {\rm cos}(\zeta_u)  \nonumber \\
Z = Im(A_{du}) & = &   \epsilon_d y_{bd} {\rm sin}(\zeta_d) - \epsilon_u y_{bu} {\rm sin}(\zeta_u)
\eeqry
then in our leading approximation for the $CKM$ matrix, the first two elements of the third column 
and row can be rewritten (making use of the phase freedom described at Eq.(\ref{eq:allCKMparams})
and immediately following) as
 \beqry
CKM13  & = &  \frac{{\rm cos}(2\omega)XY 
-\frac{1}{2}{\rm sin}(2\omega)(X^2-Y^2-Z^2)}{Q}+\frac{\imath XZ }{Q}     \\ 
CKM23   & = &  Q    \\
CKM31  & = &  \frac{{\rm cos}(\omega){\rm sin}(\omega+\Theta_C)X^2 
-{\rm sin}(\omega){\rm cos}(\omega+\Theta_C)(Y^2+Z^2)-{\rm cos}(2\omega+\Theta_C)XY}{Q}
\nonumber \\ 
& + & \frac{\imath XZ{\rm cos}(\Theta_C)}{Q}   \\
CKM32  & = & - \frac{{\rm cos}(\omega){\rm cos}(\omega+\Theta_C)X^2 
+{\rm sin}(\omega){\rm sin}(\omega+\Theta_C)(Y^2+Z^2)+{\rm sin}(2\omega+\Theta_C)XY}{Q}
\nonumber \\ 
& + & \frac{\imath XZ{\rm sin}(\Theta_C) }{Q}  \label{eq:CKMzeta}
\eeqry
where we drop the index and write $\omega$ for $\omega_u$, and where
\beqry
Q & = & \sqrt{({\rm cos}(\omega)X+{\rm sin}(\omega)Y)^2+({\rm sin}(\omega)Z)^2 } \nonumber \\
 & = & 0.04141 \label{eq:Qval}
\eeqry
and in these terms, 
\bqn
{\mathcal J} = \pm  {\rm cos}(\Theta_C)  {\rm sin}(\Theta_C) X Z 
\eqn

\begin{figure}  [h]  
\includegraphics[width=0.8\textwidth, height=0.8\textwidth, angle=0]{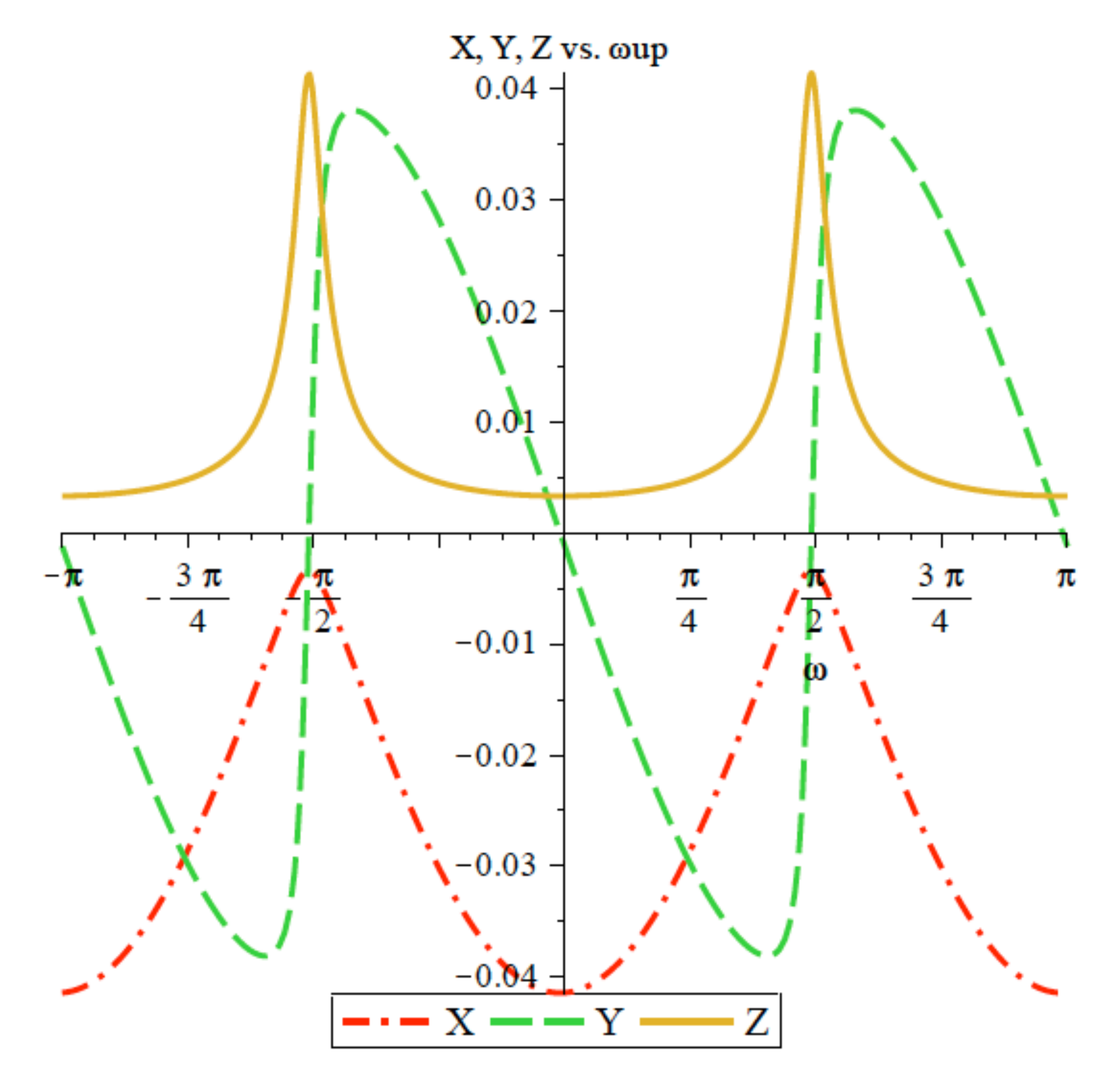}
\caption{Variation of combined BSM parameters for real and imaginary contributions 
to CKM matrix as functions of $\omega_{u}$.}
\label{fig:XYZ}
\end{figure}

It is straightforward to see from the imaginary parts that the $(3,1)$ and $(3,2)$ elements contain no 
new information beyond that from the $(1,3)$ and $(2,3)$ elements, which is true for the 
real parts also. It is also clear that the imaginary parts are consistent with the form of ${\mathcal J}$ 
as given in Eq.(\ref{eq:J2233form}). With a negative value for ${\mathcal J}$ we can solve for $X$, 
$Y$ and $Z$ as functions of $\omega_u$ using the values of ${\mathcal J}$, and of $CKM13$ and 
$CKM23$ in Eqs.(\ref{eq:CKMelvals}). We find
\beqry
X & = & \frac{- 0.00052189+0.0041169{\rm sin}(\omega_u){\rm cos}(\omega_u) - 0.070497{\rm cos}^2(\omega_u) }
{\sqrt{2.9410-2.9194{\rm sin}^2(\omega_u) -0.17049{\rm sin}(\omega_u){\rm cos}(\omega_u)}}    \\
Y & = & \frac{0.0020585 -0.070497 {\rm sin}(\omega_u){\rm cos}(\omega_u) - 0.0041169{\rm cos}^2(\omega_u) }
{\sqrt{2.9410-2.9194{\rm sin}^2(\omega_u) -0.17049{\rm sin}(\omega_u){\rm cos}(\omega_u)}}      \\
Z & = & \frac{0.005729}
{\sqrt{2.9410-2.9194{\rm sin}^2(\omega_u) -0.17049{\rm sin}(\omega_u){\rm cos}(\omega_u)}}   
\eeqry
These functions are plotted in Fig.(\ref{fig:XYZ}). Note that all quantities are no larger 
than of order $\epsilon_d$ (in fact, they are $\le 2 \epsilon_d$) and so confirm that the BSM 
parameters are consistent with being \LQ natural" in the usual sense. 

We note here that Eq.(\ref{eq:Qval}) demonstrates that Cartan sub-algebra perturbations alone 
are insufficient to satisfy these experimental constraints. If we set all of the $y_i$ to zero except 
for  $y_0, y_3$ and $y_8$, then using Eqs.(\ref{eq:ya},\ref{eq:yaomega},\ref{eq:yfval},\ref{eq:yf}), 
we have
\beqry
\frac{y_3}{\sqrt{3}} =  & y_a & = \frac{(1-\delta)}{2} {\rm sin}(2\omega) \\
\frac{y_8}{\sqrt{3}} =  & y_f  & =  \frac{(1-\delta)}{2} {\rm cos}(2\omega) 
\eeqry
or 
\beqry
{y_3} & \approx &  \frac{\sqrt{3}}{2} {\rm sin}(2\omega) \\
{y_8} & \approx &  \frac{\sqrt{3}}{2} {\rm cos}(2\omega) 
\eeqry
Under these conditions, it also follows from Eq.(\ref{eq:yc}) that $y_c = y_3 \sqrt{2/3} $. 
From Eq.(\ref{eq:yb}), we similarly obtain $y_b =y_8 \sqrt{2/3}$. Combining these in $Q$ 
(with $\zeta = 0$ so $Z = 0$ here), we find it bounded by (using the larger $\epsilon_d$)
\bqn
Q_{bound} =\epsilon \frac{ {\rm cos}(\omega)}{\sqrt{2}} < 0.02
\eqn
and so, far too small to match the value in Eq.(\ref{eq:Qval}) even if $\epsilon_u$ were to 
contribute positively. Note that using the phase freedom to produce $CP$-violation will 
not change this result. 

Conversely, it is possible for the required value of $Q$ to be attained with ${y_3} = 0$ 
or ${y_8} = 0$, and perhaps even both. Even with Eqs.(\ref{eq:constraints}) 
below satisfied, this only requires $y_5 = y_7  \sim 2$, which is still \LQ natural";  
$y_{4m6}$ provides only a small contribution. So, perhaps surprisingly, not only are 
some non-Cartan BSM contributions required, it may be that {\em both} of the Cartan 
sub-algebra symmetry-breaking components (but not $y_0$) may vanish and the 
entirety of the non-zero small masses and mixing may be due strictly to non-Cartan 
BSM contributions. 

\subsection{Inverting Definitions} \label{sec:InvDefns}
It is possible to invert the sequence of definitions of these quantities 
and return to the original $y_i$ in Eq.(\ref{eq:Mintz}). The result is 
somewhat cumbersome, but reduces the number of independent 
parameters to be considered.  Explicitly, we can set 
\beqry
y2 & = & y5 - y7 \;\;\;\;  {\rm and}  \nonumber   \\
y5 - y7  & = & 0	\label{eq:constraints}
\eeqry
These eliminate two of the imaginary terms in Eq.(\ref{eq:Mintz}) so that 
 this intermediate form of the mass matrix becomes : 
\beqry
& & {\mathcal M}_{intalt}    =   \left[ \begin{array}{ccc} 
0 & 0  & 0  \\
0 & 0  & 0  \\
0 & 0  & 1 \\
\end{array}
\right]  + \epsilon \times   \label{eq:Mintalt}    \\  
& &   \left[ \begin{array}{ccc} 
\sqrt{\frac{2}{3}} y_0 - \sqrt{\frac{1}{3}}y_8  +\frac{1}{3} ( y_1 - 2 y_{4p6}) &\frac{1}{\sqrt{3}}(- y_3 +  y_{4m6})  &
 \sqrt{\frac{2}{3}} y_8 +\frac{1}{3\sqrt{2}}( 2y_1 - y_{4p6} -\sqrt{2}  I y_{9})      \\
\frac{1}{\sqrt{3}}(- y_3 +  y_{4m6})  & \sqrt{\frac{2}{3}} y_0+ \frac{1}{\sqrt{3}}y_8 -  y_1  & 
-\frac{1}{\sqrt{6}} (2y_3 + y_{4m6}) \\
 \sqrt{\frac{2}{3}} y_8 +\frac{1}{3\sqrt{2}}( 2y_1 - y_{4p6} + \sqrt{2}  I y_{9}) & 
 -\frac{1}{\sqrt{6}} (2y_3 + y_{4m6} ) &   1+\delta-2\sqrt{\frac{2}{3}} y_0+\frac{2}{3}( y_1 +  y_{4p6}) \\
\end{array}
\right] 	\nonumber		
\eeqry
where we have replaced the labels for the equal quantities $y_5$ and $y_7$ by $y_9$.
(Recall that $\xi$ does not appear as we fixed its value by the condition that the trace of 
${\mathcal M}_{intalt}$ must be equal to the trace of the (scale removed) diagonal matrix 
 of eigenvalues.) 

Note that this procedure is not identical to the phase adjustments we made use of above. 
There the surviving real quantities in the mass matrix included contributions from the 
imaginary quantities through a normalization factor. In this truncated form, we can identify
\beqry
y_a & = &   \frac{1}{\sqrt{3}} (y_3 - y_{4m6}) 		\nonumber   \\
y_b & = & \frac{1}{3\sqrt{2}}\sqrt{ (2\sqrt{3}y_8 +2y_1-y_{4p6})^2 + 36y_9^2 }  \nonumber   \\
y_c & = & \frac{1}{\sqrt{6}} (2y_3 +y_{4m6})  \label{eq:rvsd}
\eeqry
and recall the unaltered
$
y_f =  \frac{1}{\sqrt{3}}y_8 - \frac{1}{3}(2y_1-y_{4p6}).
$

Using Eqs.(\ref{eq:ya}, \ref{eq:yfval}, \ref{eq:yaomega}, \ref{eq:yf}) and the assumptions 
in Eq.(\ref{eq:constraints}), we can simplify and invert those relations to obtain 
\beqry
2y_1 -y_{4p6} & = &  \sqrt{3}y_8 - \frac{3}{2}(1-\delta){\rm cos}(2\omega)   \label{eq:1m4p6} \\
y_{4m6}  & = & y_3 - \frac{\sqrt{3}}{2}(1-\delta){\rm sin}(2\omega)   \label{eq:4m6}
\eeqry
That is, we can rewrite these combinations of non-Cartan parameters in 
terms of $y_3$ and $y_8$ along with a dependence on $\omega$. (As 
$\omega_d = \omega_u + \Theta_C$, this is still dependence on only a
single angle, $\omega = \omega_u$.) Since only the combinations on the 
$lhs$ appear, the separate values of $y_1$, $y_4$ and $y_6$ need not 
be specified. This provides an example of how the parameters must be 
related in any BSM model. 

\section{Discussion}

The most striking result of the view espoused here is that the smallness of the 
non-Cabibbo mixing follows from the ratio of the middle to largest masses of 
the quarks. Both are due to the perturbative size of BSM corrections to the initial 
\LQ democratic" starting point. In contrast, the relatively large size of the Cabibbo 
mixing is allowed by the diagonalization process. A surprising result is that the 
BSM perturbations need not include all Cartan sub-algebra components, contrary 
to common analyses, while conversely, non-Cartan sub-algebra BSM perturbations 
{\em are} necessary. 

The mass ratios of the quarks are scale dependent, and one could examine 
the effects of that scale dependence on the CKM matrix and our fit. However, 
even the ratios are generally not that well known and do not vary significantly with scale 
over the range from 2 GeV, where the lightest quark masses are generally 
defined and determined, to the scale of the $b$-quark, nor from there to the weak 
scale which is also very close to the top quark mass. Refinements responding 
to these issues are certainly warranted, but we do not expect them to produce 
large corrections to our BSM parameter constraints determined here. In fact, 
since the effects considered here are dominated by the values of $\epsilon$, 
only the uncertainties associated with the masses of the strange and charmed 
quarks should be significant, as the $b$- and $t$-quark masses are relatively 
accurately known.  Fortunately, the very large uncertainties associated with the 
ratios of the two lightest quarks do not play a significant role in establishing the 
configuration, although they will be important for precision analyses. 

We have carried out the straightforward extension of our results to the next higher 
order in $\epsilon$ which might, in principle, be able to further constrain the values 
of the unknown parameters. Unfortunately, utilization requires knowledge of the 
relevant experimental values to order $\epsilon^2$, i.e., to of order parts in 
$10^4$, which is an accuracy generally not presently available. More accurate 
measurements would certainly change this conclusion. 

\subsection{BSM contributions}
\begin{figure} 
\includegraphics[width=0.9\textwidth, height=0.6\textwidth, angle=0]{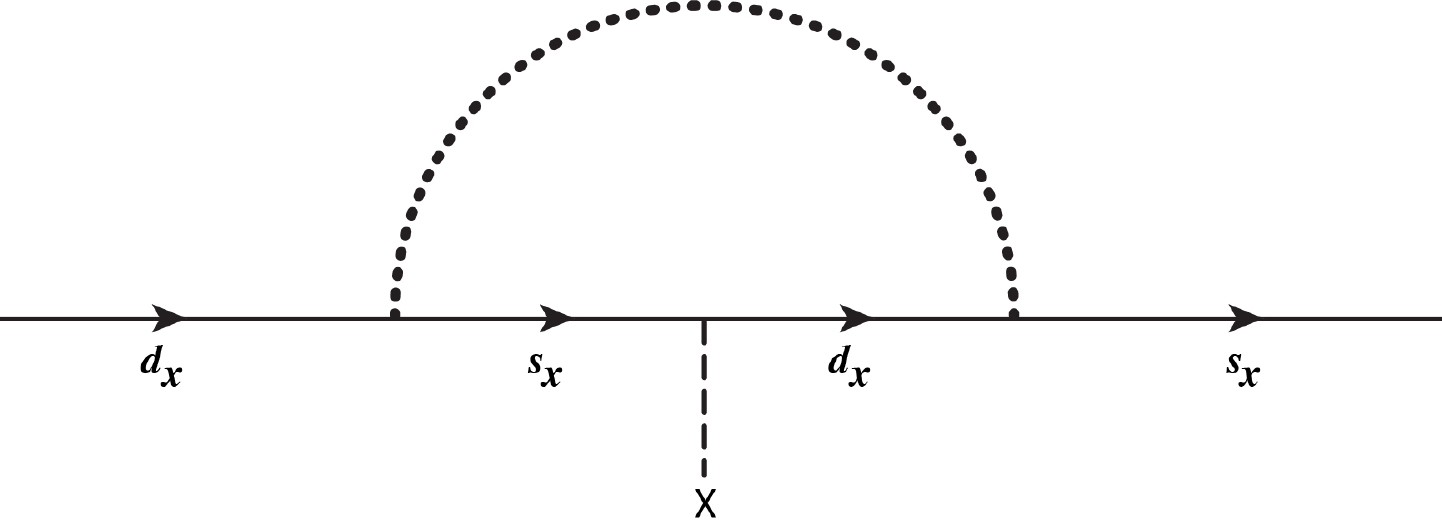}
\caption{A BSM loop correction to the fermion-Higgs-boson vertex that alters 
the charged fermion mass matrix from ``democratic" to that shown in Eq.(\ref{eq:genmtx}). 
Left-chiral fermions with weak interactions are labelled $d_x$ for \LQ doublet" and 
right-chiral fermions with no weak interactions are labelled $s_x$ for \LQ singlet" 
or \LQ sterile". }
\label{fig:chgdfermions}
\end{figure}
In general terms, Fig.(\ref{fig:chgdfermions}) shows the nature of expected BSM 
corrections that could distinguish the different fermions and lead to the small 
corrections that we find in our fits. The Lagrangian structure that we have in mind 
uses Weyl spinors for the separate left-chiral ($d_x$, for member of a weak 
interaction doublet) and right-chiral ($s_x$, for a weak interaction singlet) parts 
of the fermion Dirac bispinors, but nonetheless produces Dirac mass terms which 
may be simply represented as above. 

Interestingly, we expect these loop calculations to be finite as they involve only 
differences within the triples of fermions. This effect was observed in Ref.\cite{GG},  
where a symmetric, overall divergence appears but the differences in mass corrections 
are finite, although that model is for a quite different application of symmetry-breaking 
mass corrections and employed vector gauge bosons. 
Although the calculation there was done only for Cartan sub-algebra corrections, it  
is not unreasonable to expect that the off-diagonal corrections will also be finite. 
Note that 
without the intermediation of the Higgs scalar vacuum expectation value, the $d_x$ is 
not changed to an $s_x$ which prevents completion of the loop.

Fig.(\ref{fig:chgdfermions}) is drawn for a BSM scalar boson interaction, but a 
BSM vector could in principle also couple both the Weyl spinor $d_x$ to a  Weyl 
spinor $d_x$ and similarly $s_x$ to $s_x$. This simply requires interchanging 
the labels on either side of the Higgs' coupling to complete the loop. Again, without 
the intermediation of the Higgs scalar vacuum expectation value, both the $d_x$ 
and the $s_x$ pass through the loop unchanged without coupling to mass, so the 
BSM correction affects only vertex renormalization.

\subsection{Leptons}

Of course, application to the leptons also comes to mind. For the charged leptons, 
Fig.(\ref{fig:chgdfermions}) still applies, as it also does for Dirac mass terms for 
neutrinos.  However, producing these for the
neutrinos requires the existence of uncharged Weyl 
(right-chiral) fermions with no SM interactions at all, i.e., the so-called \LQ sterile" 
neutrinos, as opposed to the known ones  which are \LQ active" with respect to the 
weak interactions. Under the long honored assumption of quark-lepton (Zweig-Glashow) 
symmetry, the existence of these states has been widely presumed since the early 
days of Grand Unified Theories~\cite{SO10}, starting with $SO(10)$. Aside from the 
prediction of the charm quark, this symmetry (or regularity) successfully predicted 
the existence of the $t$- and $b$-quarks as soon as the $\tau$-lepton was 
discovered.~\cite{Perl} There may even be some recent experimental evidence 
for \LQ sterile" neutrinos.~\cite{MiniBooNE} 

Furthermore, with the discovery~\cite{COBE}  of \LQ Dark matter" in the Universe, 
it is clear that there are additional particles beyond those in the SM and that these 
particles are sterile. However many different types there are, only three distinct Weyl 
spinor combinations of these degrees of freedom are required to couple to the active 
neutrino Weyl spinors in conjunction with the Higgs boson to produce Dirac mass terms 
for the neutrinos. There is also no impediment for these sterile fields to acquire Majorana 
masses independently of the Higgs. 

It follows that a $6 \times 6$ structure in terms of Weyl spinors is required, rather than 
the simple $3 \times 3$ Dirac mass structure that the charged fermions can be reduced 
to by the standard construction of Dirac bispinors. (Any additional sterile fermions can 
be block-diagonalized away, leaving the three needed, albeit perhaps of a complex 
structure in terms of the original model degrees of freedom.) The Dirac mass terms of 
the neutrinos,  $m_D$, have the same form as that for the charged fermions as described 
above, but they now appear in $3 \times 3$ off-diagonal blocks in this $6 \times 6$ matrix.  
The upper left $3 \times 3$ block remains zero, as the SM does not produce Majorana 
masses for the active neutrinos, nor is it necessary for BSM physics to do so directly: The 
\LQ see-saw" mechanism~\cite{seesaw} produced by a sufficiently massive lower right 
$3 \times 3$ block of \LQ sterile" neutrinos leads to light, Majorana mass eigenstates that 
are dominated by active neutrino amplitudes. 

That lower right $3 \times 3$ block of Majorana masses (with a structure as given in 
Eq.(\ref{eq:MajM}) above) for the \LQ sterile" Weyl spinors (corresponding to what 
would have been the right-chiral component of a normal Dirac neutrino wavefunction) 
is unconstrained.  As we pointed out many years ago~\cite{CoralGables}, neutrino 
mass mixing of the  almost purely \LQ active" eigenstates can be expected to be 
similar to that for the quarks, {\em unless there is some particular structure to this 
$3 \times 3$ block of Majorana masses for the \LQ sterile" Weyl spinors}; {\it i.e.}, 
barring any special circumstances, the leptonic analog of the $CKM$ matrix,  namely 
the $PMNS$-matrix~\cite{PMNS}, should be similar to the $CKM$ matrix. 

At that time, the concern was to determine whether or not neutrinos should be expected 
to have mass and whether or not their mixing should be expected to be large enough to 
measure. As we now know, the masses are very small but the mixing is even larger than 
for the quarks and very close to the particular TBM form that we showed above applies 
separately to the up- and down-type quarks, but cancels in the $CKM$ matrix. 

We have identified a $6 \times 6$ to $3 \times 3$ block-diagonalization procedure that 
provides for a determination of the relations required between the sterile neutrino Majorana 
mass matrix and the BSM parameters in the neutrino Dirac mass matrix, so that, in the 
diagonalized $6 \times 6$ mass matrix, there is negligible mixing of the active parts of the 
mostly active Majorana mass eigenstates relative to their initial structure. That is, for the 
leptonic weak interaction currents, we have ascertained that the conditions required, so 
that the factor contributed by neutrinos to the $PMNS$-matrix is the identity (or close to it), 
may be satisfied. It follows that, under those conditions, the $PMNS$-matrix for the weak 
lepton currents will be almost, but not exactly, of the $TBM$ form with small corrections 
certainly coming from the diagonalization of the charged leptons, which may even be the 
dominant corrections: 
\bqn
PMNS \approx TBM.  \label{eq:pmns}
\eqn
This is consistent with current experiments which show that the $PMNS$-matrix is indeed 
quite close to the $TBM$-matrix form, but the quantity $\theta_{13}$, that vanishes in the 
exact limit, is not zero~\cite{dayabay}. 

If the constraints determined in our $6 \times 6$ to $3 \times 3$ block-diagonalization 
procedure are satisfied, then the \LQ democratic" plus BSM hypothesis for the fermion 
mass matrices would provide a unified understanding of all of the weak current mixing 
structures simultaneously, subject to those additional constraints on BSM physics being 
satisfied. Conversely, any model of BSM physics that satisfies these relations will produce 
the result in Eq.(\ref{eq:pmns}). We will present a detailed analysis of the leptonic sector 
in a future paper. 

\section{Conclusion}

We have started from the assumption that within the SM, for the fermions with a given electric 
charge, the Higgs doublet is sensitive only to the quantum numbers of the left-chiral Weyl 
spinor parts of the Dirac wave functions. With this assumption, a basis can be chosen so that 
the iso-singlet terms formed between them and the Higgs doublet couple equally to all of the 
right-chiral Weyl spinor parts. This in turn implies that the SM mass matrix should have a \LQ 
democratic" form, with one massive eigenstate and two massless ones. Upon adding perturbative 
corrections of a completely general form, presumed to arise from BSM physics, we find that 
a consistent set of parameters may be extracted that conforms to the known quark mass 
spectra and CKM mixing matrix, including $CP$-violation. These parameter value constraints 
provide information on matrix elements of the manner in which BSM physics couples to 
SM degrees of freedom. The question of why the overall mass scale for the up-quarks is 
significantly larger than that for the down-quarks remains unresolved, but can be accommodated 
if there are a pair of Higgs bosons, as is required in supersymmetric models, for example.

It is clear that, in this approach, extracting more detailed information on the nature of BSM 
physics and the value of BSM matrix elements  requires a more accurate determination of 
the quark mass ratios and their mixing amplitudes in the weak interaction. It would be of 
great value if the separate real and imaginary parts of the CKM matrix elements could be 
determined experimentally.

Under the \LQ see-saw" assumption regarding the existence and nature of sterile neutrino 
components, the extension of these ideas to leptons can also produce the beginning of an 
understanding as to both why the PMNS matrix is approximately of tri-bi-maximal form and 
also why it is not exactly so. Along with the information from the violations of \LQ democracy", 
the potential information on the sterile neutrino mass matrix may open the door to learning 
about the physics in the dark matter sector, with the sterile neutrinos as the first component 
known from other than gravitational interactions. 

Finally, we note that the intermediate propagation of sterile as well as active neutrinos in 
variations of the graphs above applied to weak decay box graphs may be relevant to 
recent observations of violations of lepton universality, such as in Ref.(\cite{TestLeptUniv}).

\section{Acknowledgments}
This work was carried out in part under the auspices of the National Nuclear Security
Administration of the U.S. Department of Energy at Los Alamos National Laboratory
under Contract No. DE-AC52-06NA25396. We thank Bill Louis, Geoff Mills, Richard Van 
de Water, Dharam Ahluwalia, Alan Kostele{\' c}ky, Earle Lomon, Rouzbeh Allahverdi, 
Kevin Cahill, Ami Leviatan and Xerxes Tata for useful conversations.

\end{document}